\documentclass[journal]{IEEEtran}
\usepackage{graphicx}
\usepackage{amsmath}
\usepackage{booktabs}
\usepackage{multirow}
\usepackage{graphicx}
\usepackage{booktabs}
\usepackage{amsmath}
\usepackage{amssymb}
\usepackage{algorithm}
\usepackage{algpseudocode}
\usepackage{multirow}
\usepackage[table]{xcolor}
\usepackage{graphicx}
\usepackage[caption=false,font=footnotesize]{subfig}
\usepackage{color}
\usepackage[colorlinks,
        linkcolor=blue,
        anchorcolor=blue,
        citecolor=blue]{hyperref}
\ifCLASSINFOpdf
\else
\fi
\begin{document}
%
\title{OSCAgent: Accelerating the Discovery of Organic Solar Cells with LLM Agents}
%
%
%

\author{Zhaolin Hu,
Zhiliang Wu, \emph{Member, IEEE},
Kun Li, \emph{Member, IEEE},
Hehe Fan, \emph{Senior Member, IEEE},\\
Yi Yang, \emph{Fellow, IEEE}
\thanks{Manuscript received July 30, 2026. This work was supported by the National Science and Technology Major Project (2023ZD0120803), the National Natural Science Foundation of China (62472381) and the Earth System Big Data Platform of the School of Earth Sciences, Zhejiang University. \emph{(Corresponding author: Yi Yang)}}
\thanks{Zhaolin Hu, Hehe Fan, and Yi Yang are with School of Artificial Intelligence, Zhejiang University, Hangzhou 310007, China (e-mail: 12321165@zju.edu.cn, hehefan@zju.edu.cn, and yangyics@zju.edu.cn)}
\thanks{Zhiliang Wu is with College of Computing and Data Science, Nanyang Technological University, Singapore 639798, Singapore (e-mail: zhiliang.wu@ntu.edu.sg)}
\thanks{Kun Li is with the College of Information Technology, United Arab Emirates University, Al Ain, Abu Dhabi, United Arab Emirates. (email: kun.li@uaeu.ac.ae)}
}

%
%

\markboth{IEEE Transactions on Knowledge and Data Engineering,~Vol.~, No.~, August~2026}%
{Shell \MakeLowercase{\textit{et al.}}: Bare Demo of IEEEtran.cls for IEEE Journals}
%



\maketitle

\begin{abstract}

Organic solar cells (OSCs) hold great promise for sustainable energy, but discovering high-performance materials is time-consuming and costly. Existing molecular generation methods can aid the design of OSC molecules, but they are mostly confined to optimizing known backbones and lack effective use of domain-specific chemical knowledge, often leading to unrealistic candidates. In this paper, we introduce OSCAgent, a multi-agent framework for OSC molecular discovery that unifies retrieval-augmented design, molecular generation, and systematic evaluation into a continuously improving pipeline, without requiring additional human intervention. OSCAgent comprises three collaborative agents. The Planner retrieves knowledge from literature-curated molecules and prior candidates to guide design directions. The Generator proposes new OSC acceptors aligned with these plans. The Experimenter performs comprehensive evaluation of candidate molecules and provides feedback for refinement. 
Furthermore, we introduce a multi-modal power conversion efficiency (PCE) predictor equipped with uncertainty quantification, enabling the effective integration of complementary chemical information across modalities and more reliable assessment of candidate molecules.
Experiments show that OSCAgent produces chemically valid, synthetically accessible OSC molecules and achieves superior predicted performance compared to both traditional and large language model (LLM)-only baselines. Representative results demonstrate that some candidates achieve predicted efficiencies approaching 18\%. The code will be publicly available.
\end{abstract}


%
\IEEEpeerreviewmaketitle

\section{Introduction}
As the global demand for renewable energy continues to grow, organic
solar cells (OSCs) have garnered significant attention for their
ability to efficiently convert sunlight into electricity
~\cite{wang2016difluorobenzothiadiazole,chen2019organic}. Beyond their
photovoltaic performance, OSCs offer several distinctive advantages,
including mechanical flexibility, lightweight and solution-processable
fabrication, tunable optoelectronic properties, and potentially low
manufacturing costs, making them promising alternatives to conventional
silicon-based solar technologies~\cite{dennler2009polymer}.

Despite these advantages, progress in OSCs has been substantially constrained by a continued reliance on trial-and-error discovery strategies. The development of new OSC materials typically involves complex multi-step synthesis and costly experimental characterization, rendering the exploration of chemical space both time-consuming and resource-intensive~\cite{sun2019use}. Under this paradigm, the identification of high-performance OSC candidates remains inefficient, and systematic, generalizable design principles are still lacking. Consequently, exploring data-driven approaches for OSC molecular discovery is both promising and imperative.

Recently, artificial intelligence has been extensively applied across various scientific disciplines, including biology~\cite{jumper2021highly,corso2022diffdock,wang2024protchatgpt}, physics~\cite{li2020fourier,wu2024compositional,yue2025deltaphi}, and chemistry~\cite{baum2021artificial,zhou2023uni,jian2025reaction}. Building on these advances, research on OSCs has also begun to embrace artificial intelligence as a key enabler of progress~\cite{sun2019use}. 
Most existing efforts~\cite{sahu2018toward,wu2020machine,nagasawa2018computer,meftahi2020machine,nguyen2024glad,ding2025ringformer}  have concentrated on applying machine learning models to predict molecular properties, thereby assisting in the screening and analysis of candidate materials. 
Despite promising results, the discovery of new molecules still largely relies on the expertise and intuition of domain scientists.
To alleviate this challenge, some researchers attempt to design molecules using fragment recombination, variational autoencoders (VAEs)~\cite{sun2024accelerating}, or genetic algorithms~\cite{cao2025molecular,greenstein2023screening}. 
However, these methods typically focus on optimizing the backbones of known molecules and lack the ability to effectively integrate expert chemical knowledge, failing to generate valid or diverse OSC candidates. 
Furthermore, experimentally verified high-performance OSC molecules are scarce. This scarcity makes it difficult to train robust and generalizable generative models. More importantly, existing OSC discovery methods typically treat property 
prediction, molecular generation, and candidate evaluation as separate stages. 
Consequently, evaluation outcomes cannot be systematically translated into 
updated design strategies, and promising structures identified during previous 
exploration are rarely reused to guide subsequent generations. This fragmented 
workflow limits the ability of current methods to iteratively improve candidate 
quality.

Recent advances in large language models (LLMs) provide a new
opportunity for OSC molecular discovery. By representing OSC acceptors
as chemical sequences such as SMILES, LLMs can exploit learned chemical
knowledge and in-context reasoning to analyze structural motifs,
compare known high-performance acceptors, and propose new molecular
candidates. Recent methods such as MolReGPT and ICMA further show that
retrieving relevant molecular examples can improve molecular
understanding and generation by general-purpose LLMs
~\cite{10516270,li2025large}.

In this work, we propose OSCAgent, a multi-agent framework for discovery of OSC acceptor molecules, without requiring additional human intervention. By leveraging the knowledge of LLMs in materials science, OSCAgent can explore broader regions of chemical space and generate novel candidates beyond existing structures. The framework consists of three agents. The Planner employs a retrieval-augmented~\cite{lewis2020retrieval} strategy, retrieving diverse, experimentally confirmed high-performance molecules together with dynamically updated top candidates to ground its reasoning and guide design directions. The Generator follows these directions to propose novel candidate molecules, while the Experimenter performs comprehensive evaluation using cheminformatics and machine learning tools. Candidate molecules are assessed not only for predicted power conversion efficiency (PCE)~\cite{scharber2006design}, but also for synthetic accessibility~\cite{ertl2009estimation} and electronic properties. Based on these results, the candidate database is dynamically updated to inform subsequent designs. By integrating these components, OSCAgent creates a continuously learning environment that enables the system to adapt and improve over time.


In summary, our main contributions are as follows:
\begin{itemize}
\item We introduce OSCAgent, an LLM-driven multi-agent framework for OSC
acceptor discovery, demonstrating its effectiveness in generating
chemically valid and high-performing candidate molecules.  
\item We introduce a multi-modal PCE predictor equipped with uncertainty quantification, enabling the effective integration of complementary chemical information across modalities, more reliable assessment of candidate molecules, and comprehensive feedback.
\item We introduce a retrieval-augmented design strategy that integrates diverse, validated molecules with dynamically updated top candidates, enabling informed exploration that balances novelty and feasibility.

\end{itemize}

\section{Related Work}

\subsection{Artificial Intelligence for OSCs}
Artificial intelligence has recently emerged as a powerful tool for OSCs, particularly in predicting PCE and guiding molecular design. Early studies primarily relied on handcrafted molecular fingerprints, using traditional machine learning algorithms to predict the PCE of OSC materials~\cite{mahmood2021machine,zhao2020effect}. Inspired by the success of graph neural networks (GNNs) in molecular property prediction, Eibeck et al.~\cite{eibeck2021predicting} explored their application to OSC property prediction. Building on this idea, GLaD~\cite{nguyen2024glad} adopts a multimodal approach that integrates molecular graph representations with textual descriptors to further improve PCE prediction. More recently, RingFormer~\cite{ding2025ringformer} enhances the ability of GNNs to capture ring-specific features, enabling more accurate modeling of the structural motifs that are critical in OSC molecules. Wang et al.~\cite{wang2023efficient} combined a 
GNN-based molecular property predictor with an ensemble learning model 
to directly estimate device efficiency from molecular structures. Siddiqui et al.~\cite{siddiqui2024interpretable} employed interpretable machine learning 
to identify important descriptors and distinguish high efficiency 
donor-acceptor combinations. 

Beyond predictive modeling, artificial intelligence has also been applied to OSC molecular design. Cao et al.~\cite{cao2025molecular} combined machine learning with genetic algorithms to enable efficient OSC molecular design, while DeepAcceptor~\cite{sun2024accelerating} employed a 
VAE~\cite{kingma2019introduction} framework to generate novel OSC candidate molecules. Our work introduces an LLM-based multi-agent framework that enables autonomous discovery of novel OSC acceptors.

\subsection{LLM Agents for Science} 
Large language model (LLM) agents have recently found broad applications across scientific domains, including biology~\cite{li2025intelligent, roohani2024biodiscoveryagent,su2025biomaster, ghafarollahi2024protagents,you2025developing}, chemistry~\cite{tang2025chemagent, m2024augmenting, ruan2024accelerated}, materials science~\cite{zhang2024honeycomb,hu2025electromagnetic,tian2025multi}, and physics~\cite{wuwu2025pinnsagent,liu2024physics}. PhenoGraph~\cite{niyakan2025phenograph} leverages LLM agents to automate spatial transcriptomics data analysis. ChemAgent~\cite{tang2025chemagent} enhances chemical reasoning by incorporating diverse types and functionalities of memory. OSDA Agent~\cite{hu2025osda} focuses on zeolite synthesis, integrating molecular generation, quantum evaluation, and feedback to identify suitable organic structure directing agents. PINNsAgent~\cite{wuwu2025pinnsagent} develops an automated framework for tuning physics-informed neural networks, significantly improving search efficiency. MedAgents
~\cite{tang2024medagents} coordinates agents representing different
medical specialties and uses multi-round discussion to support
collaborative clinical reasoning. EarthLink
~\cite{guo2025earthlink} integrates research planning, code execution,
climate-data analysis, and physical reasoning into a unified agentic
workflow.

Inspired by these advances, we extend the agentic framework to the autonomous discovery of high-performance OSC acceptor molecules. OSCAgent adapts the general principles of retrieval-guided planning, iterative feedback, and chemistry-aware generation to the challenges of discovering high-performance OSC acceptor molecules, thereby positioning our system as a domain-specialized instance of LLM-based scientific agents.

\begin{figure*}[t]
\centering
\includegraphics[width=0.9\textwidth]{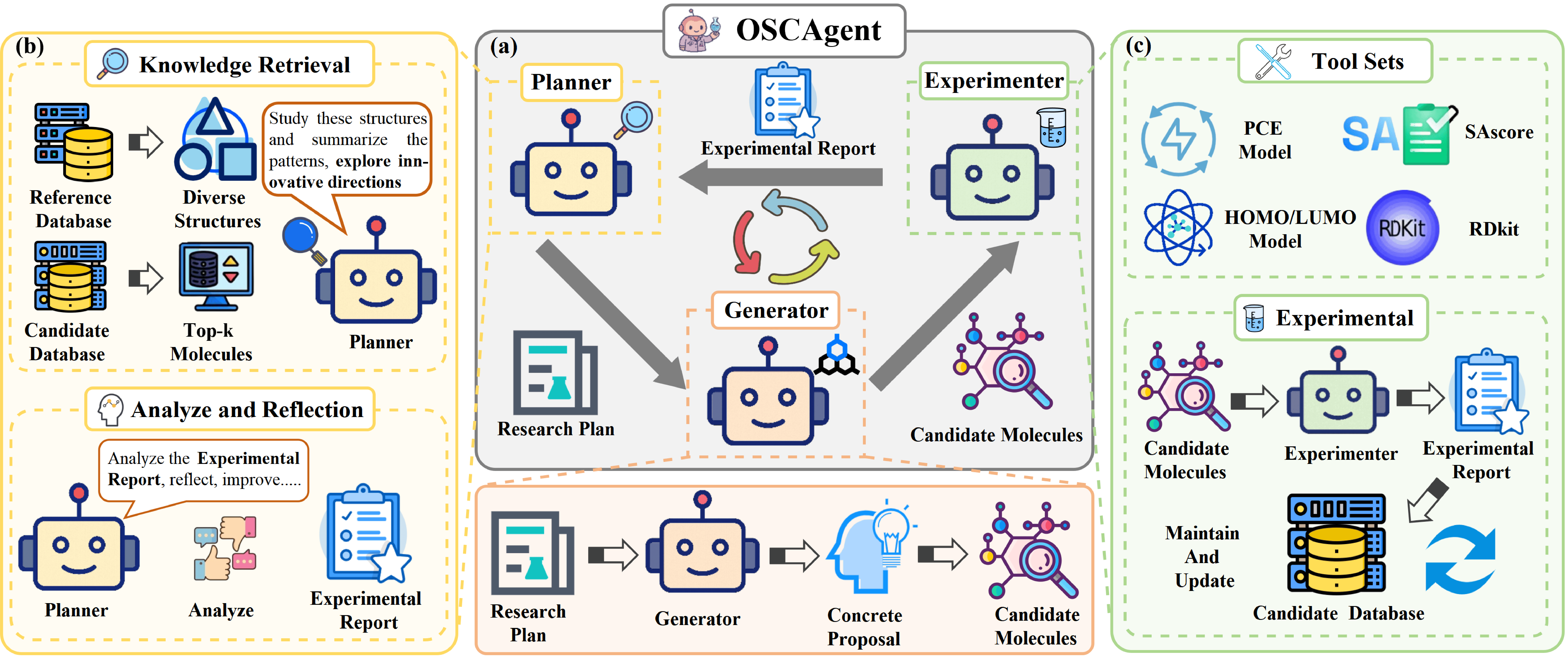} 
\caption{OSCAgent framework. (a) The pipeline consists of three collaborative agents: Planner, Generator, and Experimenter. (b) The Planner retrieves knowledge from the Reference and Candidate databases to analyze diverse structures,  and propose research plans. It further reflects on evaluation reports to refine subsequent directions. (c) The Experimenter conducts an evaluation using tools and maintains the Candidate database by incorporating promising molecules for future iterations.  }
\label{figframwork}
 \vspace{-0.3cm}
\end{figure*}

\section{Preliminaries and Data Description}

\subsection{Organic Solar Cells (OSCs)}
OSCs are a class of photovoltaic devices that use organic molecules or polymers as active materials to convert sunlight into electricity. Their operation is based on a donor–acceptor structure: the donor absorbs light and generates electron–hole pairs, which are separated at the donor–acceptor interface, with electrons transported through the acceptor material and holes through the donor material. The molecular structures of OSC components determine key properties such as light absorption and charge transport, which directly influence the overall PCE~\cite{scharber2006design}. Recent advances in non-fullerene acceptors have significantly improved OSC efficiency~\cite{zhu2021progress}, and in line with previous work~\cite{sun2024accelerating}, our study focuses on the design of OSC acceptor molecules.

\subsection{Power Conversion Efficiency (PCE)}
\label{sec32}
PCE~\cite{scharber2006design} is the key metric for assessing the performance of OSCs, measuring the proportion of incident solar energy converted into electrical power. 
PCE is defined as
\begin{equation*}
\mathrm{PCE} = \frac{V_{\mathrm{OC}} \times J_{\mathrm{SC}} \times \mathrm{FF}}{P_{\mathrm{in}}} \times 100\%,
\end{equation*}
where $V_{\mathrm{OC}}$ is the open-circuit voltage, $J_{\mathrm{SC}}$ is the short-circuit current density, $\mathrm{FF}$ is the fill factor, and $P_{\mathrm{in}}$ is the incident light power density. In practice, PCE is the principal benchmark for material screening and device optimization, and has also become a central prediction and optimization target in recent machine learning studies of OSCs~\cite{wu2020machine}. Among the factors influencing PCE, the frontier orbital energies, including the highest occupied molecular orbital (HOMO) and the lowest unoccupied molecular orbital (LUMO), play a critical role. Their relative alignment governs exciton dissociation, charge transfer, and the achievable $V_{\mathrm{OC}}$. Since HOMO and LUMO levels are strongly correlated with PCE, we incorporate them as auxiliary prediction tasks to support PCE training.

\subsection{Data}
We utilized the OSC experimental PCE dataset curated by Sun et al.~\cite{sun2024accelerating}, which contains experimentally reported OSC acceptor molecules collected from the literature. When the same molecule appeared in multiple studies, we retained the maximum reported PCE value to maintain consistency with prior work~\cite{sun2024accelerating,ding2025ringformer,sun2019machine,peng2019convolutional}. In addition, for each acceptor molecule, we calculated its synthetic accessibility score (SAscore)~\cite{ertl2009estimation}, which quantifies the estimated feasibility of synthesis on a scale from 1 (easy to synthesize) to 10 (difficult to synthesize). 

Given the scarcity of experimentally available OSC molecules, we additionally employed the computational dataset constructed by Lopez et al.~\cite{lopez2017design} as a supplement. This dataset originates from the Harvard Clean Energy Project (CEP)~\cite{hachmann2011harvard} and contains 51,256 molecules that were generated and evaluated through high-throughput density functional theory (DFT) screening to establish a curated library of potential OSC materials.

\section{Methodology}

\label{gen_inst}
\subsection{OSCAgent Architecture}
\label{sec:fm}
We propose OSCAgent, an LLM-based multi-agent framework for the discovery of acceptors for OSCs. The overall architecture is illustrated in Fig.~\ref{figframwork}. It comprises three collaborative agents, a Planner, a Generator, and an Experimenter, that coordinate to provide strategic guidance, propose candidates, and conduct comprehensive evaluations.

The Planner serves as the strategic coordinator of OSCAgent, responsible for guiding molecular exploration through a retrieval-augmented approach. It retrieves molecules from two complementary sources:  experimentally confirmed high-performance OSC molecules from the literature and a dynamically updated library of top-performing candidates (Section~\ref{sec:retrieval_strategy}). Beyond retrieval, the Planner synthesizes chemical knowledge and patterns from prior exploration to summarize insights and formulate research plans that direct the Generator. Importantly, it also reflects on the evaluation results returned by the Experimenter, revising and refining its strategies to improve subsequent design cycles. In this way, the Planner ensures that candidate generation is both scientifically grounded and adaptively responsive to experimental feedback.

The Generator operates under the Planner’s guidance, learning and understanding the research directions and translating them into concrete molecular designs and specific modification strategies. It produces candidate molecules that capture the targeted structural adjustments. Rather than engaging in unconstrained generation, the Generator emphasizes chemical plausibility and synthetic accessibility, ensuring that the proposed candidates are both scientifically meaningful and consistent with the Planner’s objectives.

The Experimenter is responsible for the systematic evaluation of candidate molecules. We establish a comprehensive assessment pipeline that covers three complementary aspects: efficiency, assessed by a multimodal PCE predictor with uncertainty quantification; synthetic accessibility, which reflects the feasibility of molecular preparation; and orbital energy consistency, evaluated through machine learning models that predict HOMO/LUMO levels to ensure candidates remain within empirically observed ranges (details in Section~\ref{sec:tools}). Based on these criteria, the Experimenter compiles structured reports that summarize candidate performance. In addition, promising molecules are added to the dynamically updated candidate library, enabling the Planner to build on high-performing structures in subsequent design iterations.

By integrating retrieval-augmented planning, guided molecular generation, and comprehensive evaluation, OSCAgent establishes a closed-loop workflow for the systematic exploration of chemical space. Through multi-agent collaboration, the framework continuously
adapts its design strategies, enabling the efficient discovery of
chemically valid and high-performing OSC acceptor candidates. A representative run of OSCAgent is shown in Fig.~\ref{figframwork2}.

\begin{figure*}[t]
\centering
\includegraphics[width=0.9\textwidth]{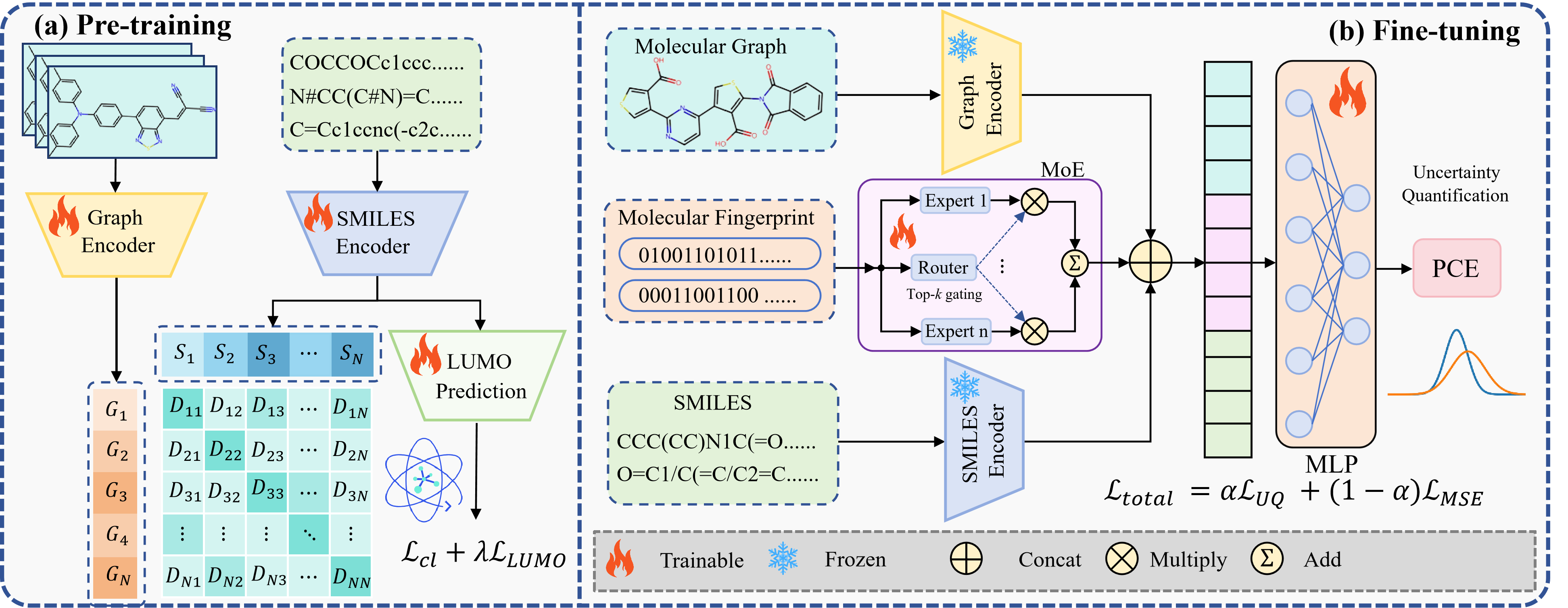} 
\caption{Details of the PCE prediction model.  
(a) Pre-training: Graph and SMILES encoders are jointly trained with contrastive learning and an auxiliary LUMO prediction task. To simplify the illustration, the LUMO prediction module of the graph branch is omitted from the figure. 
(b) Fine-tuning: Graph and SMILES embeddings are fused with molecular fingerprints via an MoE encoder, and the model is trained for PCE prediction with uncertainty quantification.      }
\label{figframwork2old}
\end{figure*}
\subsection{Comprehensive Evaluation Framework for OSC Molecular Discovery}
\label{sec:tools}
We establish a comprehensive evaluation framework to systematically assess candidate molecules along three dimensions: efficiency, synthetic accessibility, and physical feasibility. Section~\ref{sec:pce} introduces the core of this framework, the PCE prediction tool, while Section~\ref{sec:other} presents the other complementary tools.

\subsubsection{Uncertainty-Aware Multi-Modal PCE Prediction}
\label{sec:pce}

PCE is the most critical metric for evaluating the performance of
OSCs, and thus accurate prediction is essential for guiding molecular
discovery. Previous methods have primarily relied on training PCE
prediction models directly on experimental datasets. However, this
approach suffers from two major limitations. First, the scarcity of
experimental data leads to poor model performance when trained from
scratch. Second, because PCE values are experimentally measured, they
are subject to variability across laboratories, experimental
conditions, and fabrication processes, making PCE prediction
inherently uncertain.

To fully exploit the available training data, the proposed predictor
(see Fig.~\ref{figframwork2old}) is first pretrained on the large-scale
Lopez dataset~\cite{lopez2017design}. We formulate PCE prediction as a
multimodal molecular representation learning problem rather than
relying on a single input modality. The training procedure consists
of two stages. During pretraining, graph and SMILES encoders are
jointly optimized through multimodal contrastive learning and an
auxiliary LUMO prediction task. During fine-tuning, the pretrained
representations are combined with handcrafted molecular fingerprints
and used for PCE prediction.

\paragraph{Multimodal Contrastive Pretraining}
Molecular graphs and SMILES strings describe the same molecule from
complementary perspectives. Molecular graphs explicitly represent
atoms, bonds, and local chemical environments, whereas SMILES strings
encode molecular structures through sequential chemical syntax. To
align the two modalities, we jointly train a graph encoder and a
SMILES encoder using a symmetric contrastive objective.

For molecule $m_i$, let $\mathbf{z}^{G}_i$ and $\mathbf{z}^{S}_i$
denote its graph-based and SMILES-based representations, respectively.
The pair $(\mathbf{z}^{G}_i,\mathbf{z}^{S}_i)$ is treated as a positive
pair because both representations correspond to the same molecule.
Representations belonging to different molecules are treated as
negative pairs. The contrastive loss is defined as

\begin{equation}
\mathcal{L}_{\mathrm{cl}}
=
\frac{1}{2}
\left(
\mathcal{L}_{G\rightarrow S}
+
\mathcal{L}_{S\rightarrow G}
\right),
\label{eq:contrastive_loss}
\end{equation}
where the directional contrastive loss is defined as
\begin{equation}
\mathcal{L}_{A\rightarrow B}
=
-\frac{1}{N}
\sum_{i=1}^{N}
\log
\frac{
\exp\left(
\operatorname{sim}
\left(
\mathbf{z}^{A}_{i},
\mathbf{z}^{B}_{i}
\right)/\tau
\right)
}{
\displaystyle
\sum_{j=1}^{N}
\exp\left(
\operatorname{sim}
\left(
\mathbf{z}^{A}_{i},
\mathbf{z}^{B}_{j}
\right)/\tau
\right)
},
\label{eq:directional_contrastive_loss}
\end{equation}
where $(A,B)\in\{(G,S),(S,G)\}$, $N$ is the batch size,
$\tau$ is a learnable temperature parameter, and
$\operatorname{sim}(\cdot,\cdot)$ denotes the inner product between
$\ell_2$-normalized representations. Here, $G$ and $S$ denote the
molecular graph and SMILES modalities, respectively. The two
directional terms perform graph-to-SMILES and SMILES-to-graph
alignment. This symmetric objective brings representations of the same
molecule closer in the shared embedding space while separating those
of different molecules.

\paragraph{Auxiliary LUMO Prediction}
Frontier orbital energies, particularly the LUMO level, are closely
related to the electronic properties and photovoltaic performance of
OSC acceptors~\cite{scharber2006design}. To incorporate this physicochemical information into
the learned representations, we introduce an auxiliary regression
task that predicts the LUMO energy from both the graph and SMILES
branches.

Given $\mathbf{z}^{G}_i$ and $\mathbf{z}^{S}_i$, two regression heads
produce the corresponding LUMO predictions
$\hat{\ell}^{G}_i$ and $\hat{\ell}^{S}_i$. The auxiliary loss is
defined as
\begin{equation}
\mathcal{L}_{\mathrm{LUMO}}
=
\frac{1}{N}
\sum_{i=1}^{N}
\left[
\left(
\hat{\ell}^{G}_i-\ell_i
\right)^2
+
\left(
\hat{\ell}^{S}_i-\ell_i
\right)^2
\right],
\label{eq:lumo_auxiliary_loss}
\end{equation}
where $\ell_i$ denotes the ground-truth LUMO energy of molecule $i$.
This physically grounded supervision encourages both encoders to
capture electronic-structure features that are relevant to downstream
PCE prediction.

The complete pretraining objective combines multimodal contrastive
learning with auxiliary LUMO prediction:
\begin{equation}
\mathcal{L}_{\mathrm{pretrain}}
=
\mathcal{L}_{\mathrm{cl}}
+
\lambda
\mathcal{L}_{\mathrm{LUMO}},
\label{eq:pce_pretraining_loss}
\end{equation}
where $\lambda$ controls the relative contribution of the auxiliary
LUMO prediction task. The Lopez dataset provides large-scale
DFT-derived LUMO labels, enabling the model to learn transferable and
physically meaningful molecular representations before fine-tuning on
the substantially smaller experimental PCE dataset.

\paragraph{Multimodal Fine-Tuning}
During fine-tuning on limited experimental data, we additionally
incorporate handcrafted molecular descriptors
(\emph{e.g.}, Morgan and MACCS fingerprints). These descriptors are
processed through a Mixture-of-Experts (MoE)
encoder~\cite{shazeer2017sparsely} and combined with the graph- and
SMILES-based representations.

Let $\mathbf{r}_i$ denote the fingerprint representation of molecule
$i$. The fused molecular representation is given by
\begin{equation}
\mathbf{h}_i
=
\operatorname{Concat}
\left[
\mathbf{z}^{G}_i,
\mathbf{z}^{S}_i,
\operatorname{MoE}(\mathbf{r}_i)
\right].
\label{eq:pce_multimodal_fusion}
\end{equation}
In this way, the predictor integrates information from three
complementary modalities: topological structure from molecular graphs,
sequential chemical syntax from SMILES, and predefined
substructure-level patterns from molecular fingerprints. Such
multimodal fusion allows the model to exploit complementary molecular
cues and is particularly beneficial in low-data regimes, where no
single modality is sufficient to fully characterize OSC molecules.

\paragraph{Uncertainty-Aware PCE Regression}
Given the inherent noise and variability of experimental
measurements, we explicitly incorporate uncertainty quantification
into our framework by treating PCE as a random variable rather than a
fixed scalar~\cite{kendall2017uncertainties,chen2024uncertainty}.
Specifically, the predictor outputs both a mean $\mu(x_i)$ and a
variance $\sigma^2(x_i)$ for each molecule, modeling the conditional
PCE distribution as
\begin{equation}
p(y_i\mid x_i)
=
\mathcal{N}
\left(
\mu(x_i),
\sigma^2(x_i)
\right).
\label{eq:pce_gaussian_distribution}
\end{equation}

The uncertainty-aware negative log-likelihood objective is defined as
\begin{equation}
\mathcal{L}_{\mathrm{UQ}}
=
\frac{1}{N}
\sum_{i=1}^{N}
\left[
\frac{
\left(
y_i-\mu(x_i)
\right)^2
}{
2\sigma^2(x_i)
}
+
\frac{1}{2}
\log
\sigma^2(x_i)
\right].
\label{eq:pce_uq_loss}
\end{equation}
The first term penalizes prediction errors relative to the predicted
variance, whereas the logarithmic term prevents the model from
trivially assigning excessively large uncertainty to every sample.
This formulation simultaneously improves point prediction and
calibrates the predictive variance.

We also employ the standard mean squared error objective
\begin{equation}
\mathcal{L}_{\mathrm{MSE}}
=
\frac{1}{N}
\sum_{i=1}^{N}
\left(
y_i-\mu(x_i)
\right)^2.
\label{eq:pce_mse_loss}
\end{equation}
The overall fine-tuning objective combines deterministic regression
supervision with uncertainty-aware learning:
\begin{equation}
\mathcal{L}_{\mathrm{PCE}}
=
(1-\alpha)
\mathcal{L}_{\mathrm{MSE}}
+
\alpha
\mathcal{L}_{\mathrm{UQ}},
\label{eq:pce_total_loss}
\end{equation}
where $\alpha$ controls the trade-off between prediction accuracy and
uncertainty calibration.

This probabilistic formulation enables the model to report both a PCE
estimate and its associated predictive uncertainty. The Experimenter
uses the predictive uncertainty as a feedback signal and reports it to
the Planner together with the predicted PCE. This allows the Planner to
consider both predicted performance and predictive confidence when
formulating subsequent design strategies, while treating candidates
with high predicted PCE but large uncertainty more cautiously. In this
way, uncertainty-aware feedback reduces the risk of over-reliance on
noisy surrogate estimates and guides the agent toward candidates with
more reliable predictions.

In summary, our framework combines multimodal contrastive pretraining,
physically grounded LUMO supervision, fingerprint-enhanced fine-tuning,
and uncertainty-aware regression. The resulting predictor captures
complementary structural and electronic information while providing
an explicit estimate of predictive confidence. This predictive
uncertainty is further incorporated into the closed-loop molecular
design process as a feedback signal for the LLM agents. The accuracy,
calibration, and practical effect of the uncertainty estimates are
evaluated in Section~\ref{sec:uq_calibration}.

\subsubsection{Other Evaluation Tools}
\label{sec:other}
Beyond PCE prediction, OSCAgent incorporates complementary cheminformatics and property evaluation tools to provide a holistic assessment of candidate molecules.  

\textbf{SMILES Validation.} We employ RDKit~\cite{landrum2013rdkit} to parse and sanitize SMILES strings, ensuring all generated molecules are syntactically valid and chemically feasible.  

\textbf{Synthetic Accessibility.} We compute the  SAscore~\cite{ertl2009estimation} to estimate synthesis ease, using it as a constraint to ensure proposed molecules remain practically synthesizable.

\textbf{Electronic Properties.} We train predictive models for HOMO and LUMO energy levels and use them as reference indicators, encouraging values to remain within empirically observed ranges of real OSC molecules while discouraging deviations that are too large. 

In summary, these tools complement PCE prediction by ensuring that candidate molecules are efficient, chemically valid, synthetically accessible, and aligned with key orbital energy characteristics. The Experimenter systematically records their outputs to guide subsequent decisions.

\subsection{Retrieval-Augmented Strategy for OSC Molecules}
\label{sec:retrieval_strategy}

To enhance the molecular design capability of OSCAgent, we adopt a
retrieval-augmented strategy tailored to OSC acceptor molecules.
Rather than relying solely on the generic chemical knowledge acquired
during LLM pretraining, the Planner retrieves contextual information
from two complementary sources: a static Reference Database containing
experimentally validated high-performance OSC acceptors, and a dynamic
Candidate Database containing promising molecules discovered during
previous design cycles. The former provides established chemical
knowledge, whereas the latter enables the system to reuse and refine
successful experience accumulated during prior exploration.

\subsubsection{Reference Database Construction and Retrieval}
\label{sec:reference_retrieval}

The Reference Database is constructed from experimentally reported
high-performance OSC acceptors. We standardize all molecules using
RDKit through canonical SMILES normalization, tautomer unification,
and salt removal, and exclude molecules that cannot be successfully
parsed or sanitized. Each remaining molecule $m_i$ is represented by
a 2,048-bit Morgan fingerprint $\mathbf{f}_i$ with radius 2. We define
the structural distance between two molecules as
\begin{equation}
d(m_i,m_j)
=
1-
\operatorname{Tanimoto}
\left(
\mathbf{f}_i,\mathbf{f}_j
\right),
\label{eq:tanimoto_distance}
\end{equation}
where a larger distance indicates greater structural dissimilarity.

Retrieving molecules solely according to their reported PCE may produce
structurally redundant examples. We therefore use a $K$-center greedy
strategy to select high-performance acceptors with broad structural
coverage. Given the Reference Database $\mathcal{R}$ and the current
selected set $\mathcal{S}$, each new molecule is selected according to
\begin{equation}
m^{\star}
=
\arg\max_{m\in\mathcal{R}\setminus\mathcal{S}}
\min_{s\in\mathcal{S}} d(m,s).
\label{eq:kcenter_retrieval}
\end{equation}
This farthest-first criterion iteratively adds the molecule that is
most dissimilar to its nearest selected representative. The complete
retrieval procedure is summarized in
Algorithm~\ref{alg:ecfp-tani-kcenter}.

\begin{algorithm}[!t]
\caption{$K$-Center Greedy Retrieval from the Reference Database}
\label{alg:ecfp-tani-kcenter}
\footnotesize
\begin{algorithmic}[1]
\State \textbf{Input:} Valid molecule set
$\mathcal{R}=\{m_i\}_{i=1}^{N}$; retrieval size $K$
\State \textbf{Output:} Selected molecule set $\mathcal{S}$

\For{each molecule $m_i\in\mathcal{R}$}
  \State
  $\mathbf{f}_i
  \gets
  \mathrm{Morgan}
  (m_i;\mathrm{radius}=2,\mathrm{nBits}=2048)$
\EndFor

\State Sample an initial molecule $m_{i_0}$ from $\mathcal{R}$
\State $\mathcal{S}\gets\{m_{i_0}\}$
\For{each $m_i\in\mathcal{R}$}
  \State
  $\Delta_i\gets d(m_i,m_{i_0})$
  \Comment{distance to the selected set}
\EndFor

\While{$|\mathcal{S}|<K$}
  \State
  $i^{\star}
  \gets
  \arg\max_{i:m_i\notin\mathcal{S}}
  \Delta_i$
  \State
  $\mathcal{S}
  \gets
  \mathcal{S}\cup\{m_{i^{\star}}\}$
  \For{each $m_i\in\mathcal{R}\setminus\mathcal{S}$}
    \State
    $\Delta_i
    \gets
    \min\!\left(
    \Delta_i,
    d(m_i,m_{i^{\star}})
    \right)$
  \EndFor
\EndWhile

\State \textbf{Return} $\mathcal{S}$
\end{algorithmic}
\end{algorithm}

In practice, we set $K_{\mathrm{ref}}=5$. For each selected molecule,
we provide the Planner with its SMILES representation and associated
properties, including the experimentally reported PCE, SA score, and
HOMO and LUMO energy levels. The resulting reference set provides
experimentally grounded structural motifs while avoiding excessive
redundancy among the retrieved examples.

\subsubsection{Candidate Database Retrieval and Maintenance}
\label{sec:candidate_retrieval}

The Candidate Database is dynamically maintained throughout the
closed-loop design process. After each design cycle, the Experimenter
evaluates the generated molecules and records promising candidates
together with their predicted PCE, synthetic accessibility score, and
HOMO/LUMO energy levels. The Candidate Database therefore accumulates
the most informative outcomes of previous exploration and allows the
system to reuse its own successful design experience.

The accumulated candidates are ranked using a composite score that
jointly considers photovoltaic efficiency, synthetic feasibility, and
frontier-orbital consistency:
\begin{equation}
\operatorname{Score}(m)
=
\widehat{\operatorname{PCE}}(m)
-
\operatorname{SAscore}(m)
+
f\!\left(h(m),l(m)\right),
\label{eq:candidate_score}
\end{equation}
where $\widehat{\operatorname{PCE}}(m)$ is the predicted PCE,
$\operatorname{SAscore}(m)$ measures the estimated synthetic
difficulty, and $h(m)$ and $l(m)$ denote the predicted HOMO and LUMO
energy levels, respectively. A high predicted PCE increases the
ranking score, whereas a high SAscore penalizes molecules that are
expected to be difficult to synthesize.

\begin{figure}[!t]
\centering
\includegraphics[width=0.85\columnwidth]
{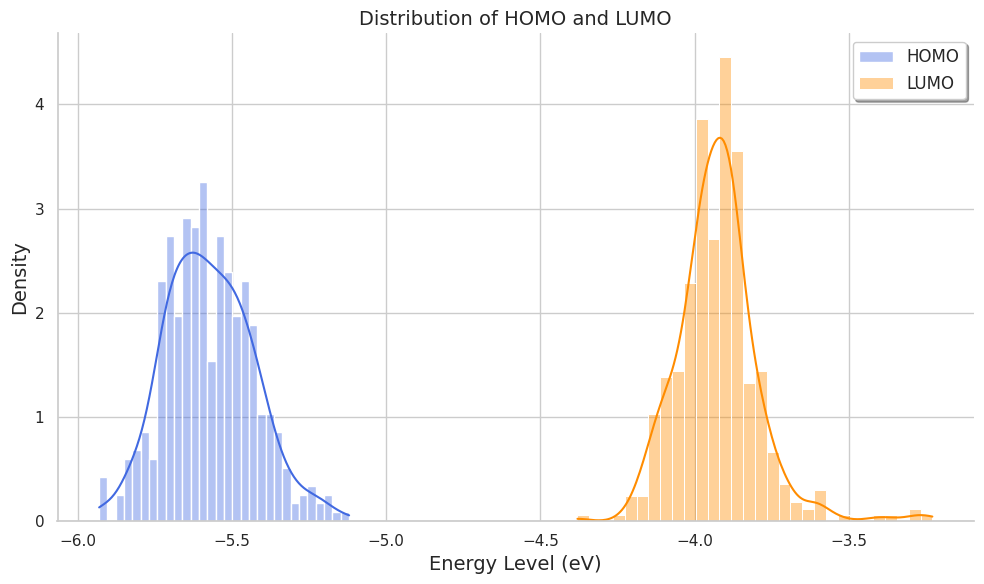}
\caption{Distributions of the HOMO and LUMO energy levels in the
literature-curated OSC acceptor dataset~\cite{sun2024accelerating}.}
\label{fig:homo_lumo_distribution}
\end{figure}

To discourage molecules with atypical frontier-orbital energies, we
introduce an interval-based feasibility adjustment. Based on the
distributions observed in the literature-curated OSC acceptor dataset,
we define the empirical HOMO and LUMO ranges as
$[-6.0,-5.0]$~eV and $[-4.5,-3.0]$~eV, respectively, as shown in
Fig.~\ref{fig:homo_lumo_distribution}. Candidates whose predicted
HOMO and LUMO values simultaneously fall within these ranges receive
a reward of 3, whereas candidates outside the ranges receive a penalty
of 3. This adjustment favors molecules with frontier-orbital energies
consistent with those commonly observed in OSC acceptors.

At each retrieval step, the highest-scoring candidates are selected from the Candidate Database. Because the database is continuously updated by the Experimenter, the retrieved examples evolve throughout the design process. These dynamic exemplars enable the Planner to build upon promising structures identified in previous iterations.

\subsubsection{Retrieval-Augmented Prompting}
\label{sec:retrieval_prompting}

At the beginning of each design cycle, the Planner retrieves
$K_{\mathrm{ref}}$ structurally diverse molecules from the static
Reference Database and $K_{\mathrm{cand}}$ top-ranked molecules from
the dynamic Candidate Database. Each retrieved molecule is represented
by its SMILES string and associated properties, including PCE,
SAscore, HOMO, and LUMO values. These examples are inserted into the
Planner's prompt as contextual knowledge.

The two databases play complementary roles. The Reference Database
grounds the Planner in experimentally validated structural motifs and
established OSC design principles, while its diversity-oriented
retrieval mechanism prevents the prompt from being dominated by highly
similar molecular structures. In contrast, the Candidate Database
provides adaptive feedback from the system's previous exploration and
highlights molecular modifications that have already demonstrated
promising predicted performance.

Based on these complementary sources, the Planner analyzes recurring
structural patterns, compares the properties of retrieved molecules,
identifies promising modification directions, and formulates a
research plan for the Generator. The Generator then translates this
plan into concrete molecular candidates. These candidates are
evaluated by the Experimenter in terms of chemical validity, predicted
PCE, synthetic accessibility, and frontier-orbital consistency.
Promising molecules are subsequently added to the Candidate Database
and become available for retrieval in later design cycles.

This process forms a continuously improving
retrieval--generation--evaluation loop. The static Reference Database
preserves experimentally grounded chemical knowledge, whereas the
dynamic Candidate Database acts as a property-guided memory of prior
exploration. By integrating these two information sources, the
retrieval-augmented strategy balances structural diversity,
photovoltaic performance, and practical synthetic feasibility. The
influence of the retrieval sizes $K_{\mathrm{ref}}$ and
$K_{\mathrm{cand}}$ is evaluated in
Section~\ref{sec:retrieval_size}.

\begin{table*}[!t]
\centering
\caption{Performance comparison of different molecular generation
methods. Best values are shown in \textbf{bold}, and the second-best
distinct values are \underline{underlined}. Multiple tied best values
are all boldfaced. All GPT-based baselines use GPT-5 as the backbone
unless otherwise specified.}
\label{tab:gen_methods}
\setlength{\tabcolsep}{5pt}
\renewcommand{\arraystretch}{1.2}

\resizebox{0.90\textwidth}{!}{
\begin{tabular}{lcccccccc}
\toprule
\multirow{2}{*}{Method}
& \multicolumn{2}{c}{Diversity}
& \multicolumn{2}{c}{Molecular Quality}
& \multicolumn{4}{c}{Distribution Similarity} \\
\cmidrule(lr){2-3}
\cmidrule(lr){4-5}
\cmidrule(lr){6-9}
& Uniqueness $\uparrow$
& Novelty $\uparrow$
& Validity $\uparrow$
& Avg.~PCE (\%) $\uparrow$
& Morgan $\uparrow$
& MACCS $\uparrow$
& RDK $\uparrow$
& ECFP6 $\uparrow$ \\
\midrule

BRICS~\cite{degen2008art}
& 0.871
& \textbf{1.000}
& 0.049
& 6.461
& 0.337
& 0.694
& 0.681
& 0.276 \\

VAE~\cite{kingma2019introduction}
& \underline{0.919}
& \textbf{1.000}
& 0.002
& 4.967
& 0.219
& 0.611
& 0.663
& 0.167 \\

SMILES-GA~\cite{brown2019guacamol}
& 0.626
& 0.866
& 0.112
& 7.275
& 0.268
& 0.597
& \underline{0.826}
& 0.239 \\

Graph-GA~\cite{jensen2019graph}
& 0.884
& 0.823
& 0.180
& 7.543
& 0.381
& 0.606
& 0.641
& 0.362 \\

BioT5~\cite{pei2023biot5}
& 0.802
& \textbf{1.000}
& 0.000
& 3.573
& 0.132
& 0.424
& 0.462
& 0.114 \\

REINVENT4~\cite{loeffler2024reinvent}
& 0.872
& \textbf{1.000}
& 0.003
& 5.590
& 0.231
& 0.570
& 0.703
& 0.189 \\

Few-shot (GPT-5)
& 0.709
& 0.989
& 0.283
& 9.249
& 0.394
& 0.693
& 0.684
& 0.326 \\

Vanilla Agent (GPT-5)
& 0.835
& \textbf{1.000}
& 0.313
& 9.870
& 0.341
& 0.712
& 0.733
& 0.275 \\

\midrule
\rowcolor{gray!10}
\textbf{OSCAgent (GPT-4o)}
& \textbf{0.937}
& \underline{0.992}
& \underline{0.628}
& \underline{13.010}
& \underline{0.436}
& \underline{0.721}
& 0.793
& \underline{0.363} \\

\rowcolor{gray!10}
\textbf{OSCAgent (GPT-5)}
& 0.893
& \textbf{1.000}
& \textbf{0.705}
& \textbf{14.590}
& \textbf{0.475}
& \textbf{0.748}
& \textbf{0.857}
& \textbf{0.395} \\

\bottomrule
\end{tabular}
}
\end{table*}

\section{Experiments}

This section presents the experimental evaluation of OSCAgent.
Section~\ref{sec:setup} introduces the experimental setup, including
the datasets, baseline methods, and implementation details.
Section~\ref{sec:results} evaluates OSC molecular design through
quantitative comparisons, ablation studies, and representative case
studies. Section~\ref{sec:pce2} reports the performance of the PCE
prediction model.

\subsection{Experimental Setup}
\label{sec:setup}
In the OSCAgent framework, the Planner, Generator, and Experimenter all invoke the GPT-5 API. The predictive models used by the Experimenter for predicting PCE, HOMO/LUMO are trained on the Lopez dataset~\cite{lopez2017design} together with the experimental dataset collected by Sun et al.~\cite{sun2024accelerating}. 


We compare OSCAgent with a diverse set of baselines covering
rule-based generation, latent-variable models, genetic algorithms,
reinforcement-learning-based molecular design, and LLM-based
approaches. For rule- and latent-space-based generation, we adopt
BRICS~\cite{degen2008art} and VAE~\cite{kingma2019introduction}
following the implementations used in
DeepAcceptor~\cite{sun2024accelerating}. BRICS decomposes existing
molecules according to predefined chemical rules and generates new
candidates by recombining the resulting fragments, whereas the VAE
encodes molecules into a continuous latent space and samples new
structures through latent-space decoding.

For evolutionary molecular optimization, we evaluate
SMILES-GA~\cite{brown2019guacamol} and
Graph-GA~\cite{jensen2019graph}. These methods iteratively optimize
candidate molecules through mutation and crossover operations applied
to SMILES strings and molecular graphs, respectively, using
OSC-related property scores as the optimization objective. We also
include REINVENT4~\cite{loeffler2024reinvent}, a reinforcement-learning-based molecular design
framework that optimizes a generative policy according to the same
property-oriented evaluation criteria used for the other generation
baselines.

For language-model-based approaches, we evaluate
BioT5~\cite{pei2023biot5}, a domain-specific text-to-molecule model
pretrained on large-scale molecular and biomedical corpora. We further
consider two general-purpose LLM baselines. In the Few-shot + Direct
Reasoning baseline, a curated set of high-performance OSC acceptors is
provided as in-context demonstrations, and the model directly proposes
new molecules without iterative agentic refinement. The Vanilla Agent
baseline employs a general-purpose agent to iteratively generate and
revise molecular candidates, but does not use the complete
retrieval-augmented Planner--Generator--Experimenter collaboration of
OSCAgent.


For a fair comparison, all general-purpose LLM baselines, including
Few-shot + Direct Reasoning and Vanilla Agent, use GPT-5 as the
backbone and share the same molecular property predictors and validity
criteria as OSCAgent. We additionally report OSCAgent results with
GPT-4o to assess the effect of the underlying LLM backbone. All
experiments are conducted on a GPU cluster equipped with NVIDIA RTX
4090, RTX A6000, and L40S GPUs.

\begin{table*}[ht]
\centering
\caption{Ablation study of OSCAgent. 
We report diversity (uniqueness), molecular quality (validity and average PCE), and distributional similarity scores computed with four fingerprints (Morgan, MACCS, RDK, ECFP6). 
Removing either the Retrieval-Augmented Strategy or the Experimenter results in a clear degradation of performance across multiple metrics.}
\label{tab:ablation}
\setlength{\tabcolsep}{5pt} 
\renewcommand{\arraystretch}{1.2} 
\resizebox{0.9\textwidth}{!}{
\begin{tabular}{lccccccc}
\toprule
\multirow{2}{*}{Method} 
& \multicolumn{1}{c}{Diversity} 
& \multicolumn{2}{c}{Molecular Quality} 
& \multicolumn{4}{c}{Distribution Similarity} \\
\cmidrule(lr){2-2} \cmidrule(lr){3-4} \cmidrule(lr){5-8}
& Uniqueness \(\uparrow\) & Validity \(\uparrow\) & Avg.~PCE(\%) \(\uparrow\) 
& Morgan \(\uparrow\) & MACCS \(\uparrow\) & RDK \(\uparrow\) & ECFP6 \(\uparrow\) \\
\midrule
OSCAgent (full) & \textbf{0.893} & \textbf{0.705} & \textbf{14.59} 
                     & \textbf{0.475} & \textbf{0.748} & \textbf{0.857} & \textbf{0.395} \\
\textit{w/o} Retrieval-Aug. Strategy & 0.813 & 0.518 & 10.41 
                     & 0.414 & 0.723 & 0.817 & 0.356 \\
\textit{w/o} Experimenter           & 0.847 & 0.387 & 13.21
                     & 0.457 & 0.735 & 0.845 & 0.370 \\
\bottomrule
\end{tabular}
}
\end{table*}

\subsection{Results of OSC Molecule Design}
\label{sec:results}
We evaluate the effectiveness of OSCAgent by comparing it with traditional molecular generation methods (BRICS and VAE) and LLM-based methods (BioT5 and Few-shot + Direct Reasoning). For fairness, the GPT-5 models in the Few-shot setting were provided with the same high-performance OSC molecules that served as prompts in OSCAgent.

To assess performance, we use eight established metrics covering diversity, effectiveness, and distributional alignment. Diversity is quantified by uniqueness, the proportion of distinct molecules among generated candidates, and novelty, the fraction of molecules absent from the existing dataset. Effectiveness is evaluated by validity, \emph{i.e.}, the proportion of chemically valid molecules that satisfy design criteria (PCE $>$ 10\%, SAscore $<$ 8), and by average PCE, the mean predicted efficiency of generated molecules.
We further evaluate the distributional similarity between generated
molecules and experimentally reported high-performance OSC acceptors
using the Sinkhorn--Wasserstein distance
~\cite{cuturi2013sinkhorn}. For a given fingerprint representation,
let $\mathcal{X}=\{\mathbf{x}_i\}_{i=1}^{n}$ and
$\mathcal{Y}=\{\mathbf{y}_j\}_{j=1}^{m}$ denote the generated and
reference molecular sets, respectively. The Sinkhorn--Wasserstein
distance is defined as
\begin{equation}
\begin{aligned}
W_{\varepsilon}(\mathcal{X},\mathcal{Y})
&=
\min_{\boldsymbol{\Gamma}\in\Pi(\mathbf{a},\mathbf{b})}
\Bigg[
\sum_{i=1}^{n}\sum_{j=1}^{m}
\Gamma_{ij}C_{ij} \\
&\qquad\qquad
+
\varepsilon
\sum_{i=1}^{n}\sum_{j=1}^{m}
\Gamma_{ij}
\left(\log\Gamma_{ij}-1\right)
\Bigg],
\end{aligned}
\label{eq:sinkhorn_wasserstein}
\end{equation}
where $C_{ij}=1-\operatorname{Tanimoto}
(\mathbf{x}_i,\mathbf{y}_j)$ is the transport cost between two
fingerprint vectors, $\boldsymbol{\Gamma}$ is the transport plan, and
$\varepsilon$ is the entropy regularization coefficient. We compute
the distance separately using Morgan, MACCS, RDK, and ECFP6
fingerprints. The corresponding distributional similarity score is
defined as
\begin{equation}
S_{\varepsilon}(\mathcal{X},\mathcal{Y})
=
1-W_{\varepsilon}(\mathcal{X},\mathcal{Y}),
\label{eq:sinkhorn_similarity}
\end{equation}
where a larger value indicates stronger distributional alignment between the generated and reference molecular sets.

\begin{figure*}[t]
\centering
\includegraphics[width=0.95\textwidth]{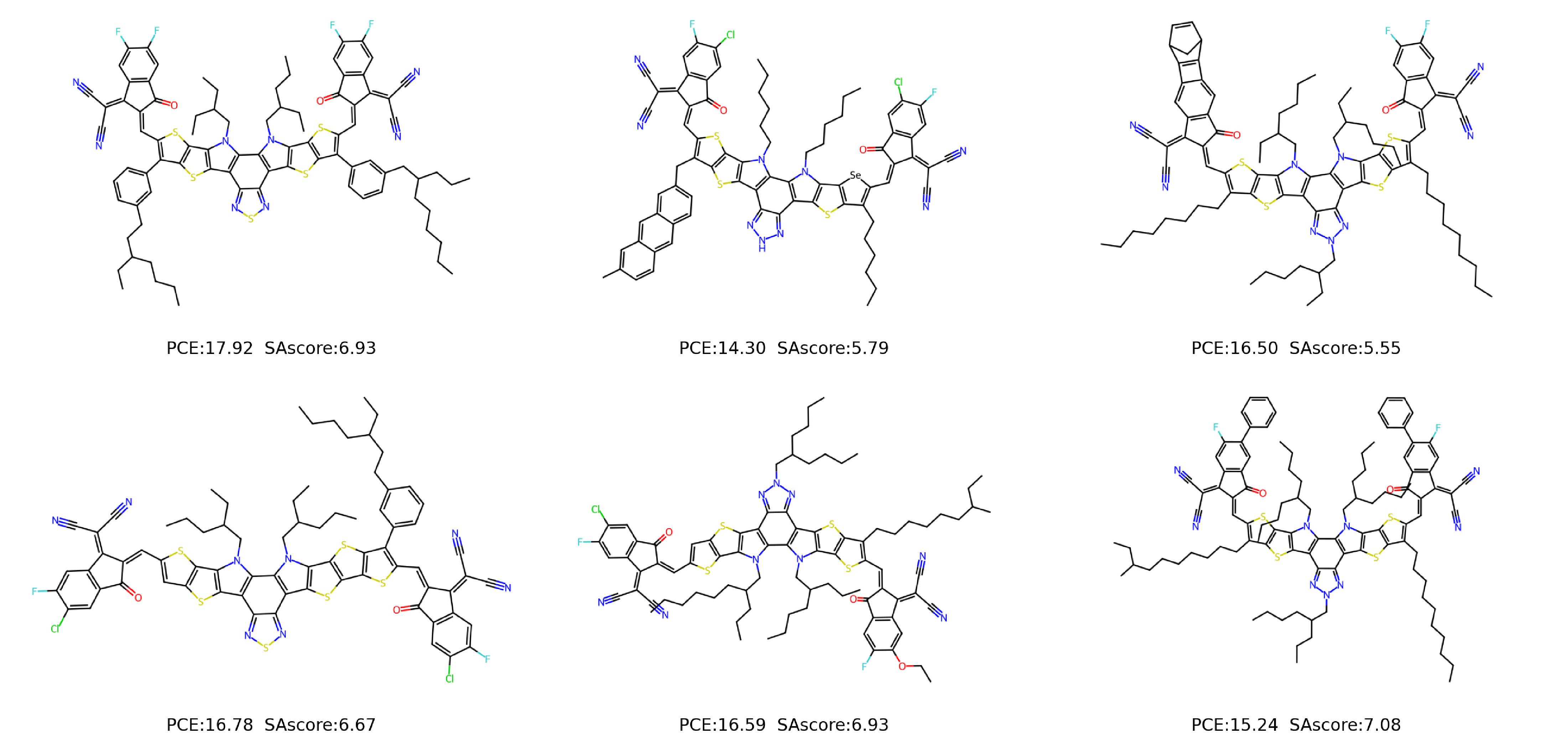} 
\caption{Representative OSC molecules generated by OSCAgent. These
candidates exhibit favorable predicted photovoltaic properties and
synthetic accessibility. Expert assessment further suggests that they
possess promising potential for future synthesis and experimental
validation.}
\label{fig:case_studies}
\end{figure*}

As shown in Table~\ref{tab:gen_methods}, traditional molecular generation approaches such as BRICS and VAE rely on fragment recombination or latent space sampling, essentially exploring variations of existing structures without clear design guidance. Consequently, most generated candidates are chemically infeasible or exhibit poor performance. Genetic algorithm–based methods, including SMILES-GA and Graph-GA, introduce iterative optimization through mutation and crossover, leading to moderate improvements in validity and average PCE; however, their search remains largely local and heuristic, limiting both diversity and overall performance. Among language model-based methods, BioT5 faces significant limitations because it lacks OSC-specific chemical knowledge, making it difficult to generate candidates that meet performance requirements. In contrast, general-purpose LLMs like GPT, guided by few-shot prompting with strong examples, can produce more reasonable molecules. Nevertheless, without the integration of specialized chemical tools, the few-shot approach still suffers from clear shortcomings in both accuracy and diversity.

Among the LLM-based baselines, the Vanilla Agent improves validity and
average PCE over direct few-shot generation, demonstrating the benefit
of iterative refinement. However, it remains substantially below
OSCAgent, indicating that generic agentic iteration alone is
insufficient. The retrieval-augmented planning mechanism, specialized
multi-agent collaboration, and systematic evaluation feedback are
necessary to consistently identify chemically valid and
high-performance OSC acceptors.

In contrast, the proposed OSCAgent demonstrates consistent superiority over all baseline methods. By combining the chemical knowledge encoded in LLMs with domain-specific evaluation tools and knowledge-augmented design strategies, OSCAgent is able to generate OSC molecules that are both chemically valid and performance-oriented. As shown in Table~\ref{tab:gen_methods}, it achieves the best validity and the highest average PCE, while also obtaining the strongest distributional similarity to real high-performance molecules across multiple fingerprinting metrics. These results highlight that OSCAgent not only produces feasible and diverse structures, but also identifies candidates with superior photovoltaic potential, underscoring the advantage of an iterative, knowledge-augmented multi-agent framework for OSC discovery. 



\subsubsection{Ablation Study}

To better understand the contribution of each component in the OSCAgent framework, we conduct ablation experiments along three dimensions.  

\paragraph{Effect of Retrieval-Augmented Strategy}
In this setting, the Planner relies solely on fixed few-shot prompts and the general knowledge acquired during LLM pretraining, without retrieving molecules from either the Reference Database of experimentally confirmed OSCs or the Candidate Database updated from prior feedback. As shown in Table~\ref{tab:ablation}, this results in a clear decline in performance, with both predicted PCE and distributional similarity scores dropping compared to the full model. These findings suggest that static prompting alone is insufficient, and that retrieval-augmented knowledge from literature-derived references and prior experience is crucial for guiding the design process toward chemically realistic and structurally diverse candidates.

\paragraph{Effect of Experimenter}  
In this setting, the Planner receives no feedback from chemical evaluation tools and must rely entirely on its own pretrained knowledge and heuristic judgment when formulating design plans. Without this feedback loop, the design process loses critical guidance: although structural diversity remains comparable, candidate validity and predicted efficiency degrade noticeably. These results underscore the pivotal role of the Experimenter in providing systematic chemical assessment and feedback, which steers exploration toward chemically feasible and high-performing regions of the design space.

\begin{figure*}[t]
\centering
\includegraphics[width=0.9\textwidth]{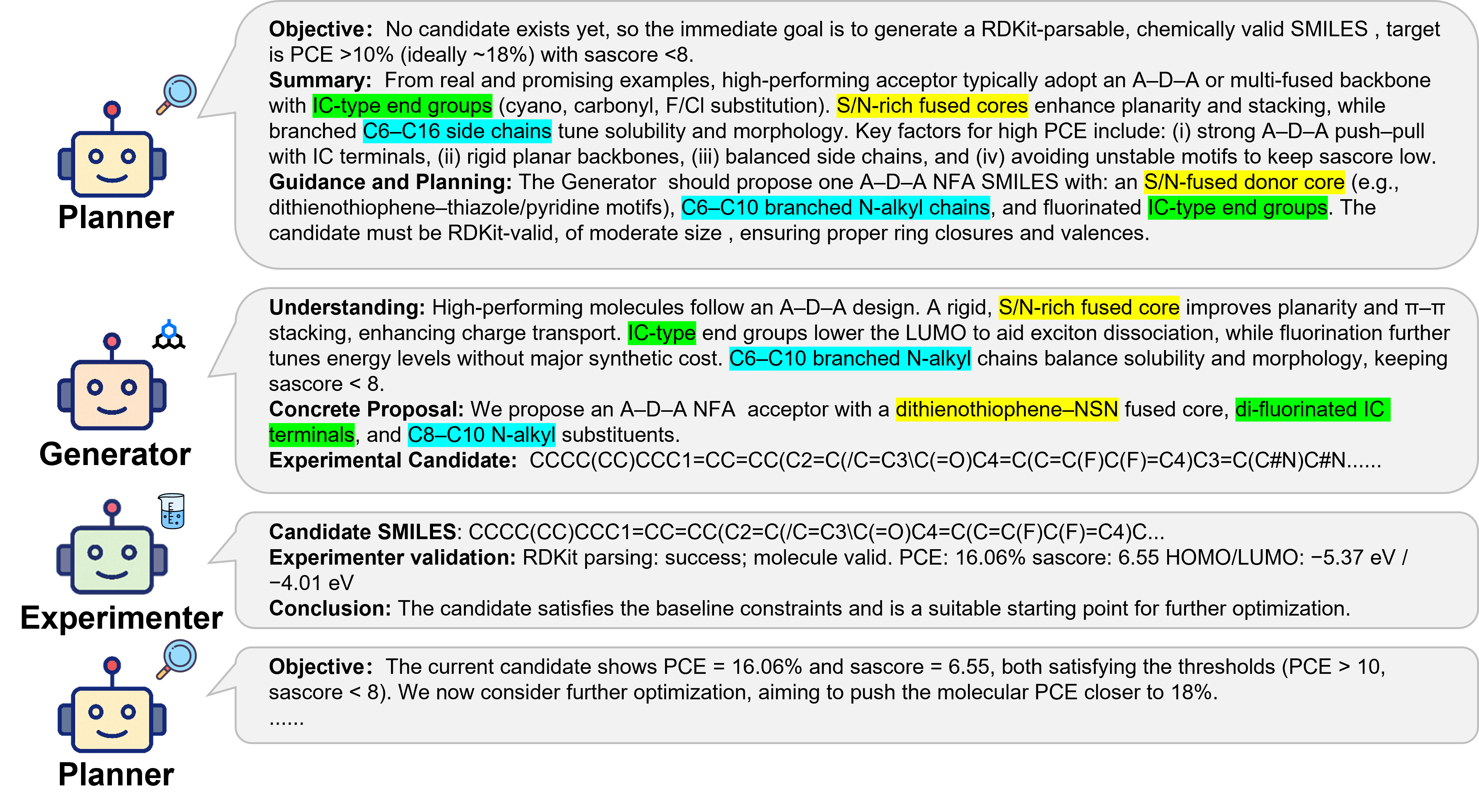} 
\caption{Example of agent collaboration within OSCAgent for OSC molecular
design. For clarity, a simplified version of the complete agent
dialogue is presented, with related information highlighted using the
same color. }
\label{figframwork2}
\end{figure*}

Overall, these results demonstrate that both the Retrieval-Augmented Strategy and the Experimenter are indispensable. The former enriches the Planner’s reasoning with literature knowledge and prior feedback, while the latter provides systematic chemical evaluation to refine candidate quality. Together, they enable OSCAgent to achieve more effective and reliable molecular discovery. 

\paragraph{Effect of Retrieval Size}
\label{sec:retrieval_size}

We further examine how the number of molecules retrieved from the Reference and Candidate databases affects molecular generation. We denote the two retrieval sizes as $(K_{\mathrm{ref}},K_{\mathrm{cand}})$. 
As shown in Table~\ref{tab:retrieval_size}, using only two molecules from each database provides insufficient structural and property diversity, leading to consistently lower distributional similarity. Increasing both retrieval sizes to seven does not produce consistent improvements
over the default configuration. In particular, the larger prompts
substantially increase token consumption because OSC acceptors
typically contain long and structurally complex SMILES strings.

We therefore use $K_{\mathrm{ref}}=5$ and
$K_{\mathrm{cand}}=3$, which provides a favorable balance between
structural coverage and computational efficiency. A larger retrieval
size is assigned to the Reference Database because experimentally
validated molecules provide more reliable chemical anchors, whereas
the Candidate Database mainly supplies adaptive feedback from previous
iterations.

\begin{table}[t]
\centering
\caption{Effect of the Reference and Candidate database retrieval
sizes, denoted by $(K_{\mathrm{ref}},K_{\mathrm{cand}})$, on the
distributional similarity of generated molecules.}
\label{tab:retrieval_size}
\setlength{\tabcolsep}{4pt}
\renewcommand{\arraystretch}{1.08}
\resizebox{0.8\columnwidth}{!}{
\begin{tabular}{ccccc}
\toprule
$(K_{\mathrm{ref}},K_{\mathrm{cand}})$
& Morgan $\uparrow$
& MACCS $\uparrow$
& RDK $\uparrow$
& ECFP6 $\uparrow$ \\
\midrule
$(2,2)$
& 0.422
& 0.719
& 0.821
& 0.331 \\
$(5,3)$
& \textbf{0.475}
& \textbf{0.748}
& 0.857
& \textbf{0.395} \\
$(7,7)$
& 0.457
& 0.747
& \textbf{0.859}
& 0.379 \\
\bottomrule
\end{tabular}
}
\end{table}

\subsubsection{Case Studies }
To qualitatively illustrate the design capability of OSCAgent, we present two representative examples.  

\paragraph{Examples of Representative OSC Molecules}
Using OSCAgent, we generated nearly 1,000 OSC acceptor molecules with predicted PCE values above 14\%. To qualitatively demonstrate the molecular design capability of the proposed framework, representative candidates are presented in Fig.~\ref{fig:case_studies}. These molecules retain hallmark structural characteristics of high-performance OSC acceptors. All representative candidates achieve predicted PCE values exceeding 14\% and SAscores below 8.0, indicating a favorable balance between photovoltaic performance and synthetic feasibility. Expert assessment further suggests that these molecules possess promising potential for high-performance OSC applications. Collectively, these case studies demonstrate the ability of OSCAgent to balance molecular novelty, synthetic practicality, and photovoltaic performance.

\paragraph{Example of OSCAgent's Multi-Agent Collaboration}
We provide a representative case study to illustrate the collaborative process of our LLM-driven agent system for OSC molecular design. As shown in Fig.~\ref{figframwork2}, the Planner first retrieves knowledge from the Reference Database and previous candidate molecules, identifying useful design patterns such as the common A--D--A framework for high-performance OSC acceptors, with S/N-rich fused cores, IC-type end groups, and balanced side chains. Based on this information, the Planner formulates a high-level design plan for the Generator.

Following this plan, the Generator turns these design principles into a concrete molecular proposal. In this example, it suggests a structure with a dithienothiophene--NSN fused core, di-fluorinated IC terminals, and C8--C10 N-alkyl side chains. The proposed molecule is then passed to the Experimenter, which evaluates it using key metrics such as predicted PCE, synthetic accessibility, and HOMO/LUMO values.

Through multiple iterations, the three agents play complementary roles: the Planner defines the design direction, the Generator proposes chemically valid candidates, and the Experimenter provides quantitative feedback. This case study shows how the multi-agent framework integrates knowledge retrieval, molecular generation, and evaluation into a closed-loop process for OSC molecular design.
\subsection{Results of PCE Prediction}
\label{sec:pce2}

To assess the effectiveness of our PCE predictor, we evaluate it on the OSC experimental dataset~\cite{sun2024accelerating}, comparing with baseline models and ablation studies on key components. Table~\ref{tab:pce_results} reports the performance in terms of $R^2$ and MAE.

For baselines, we consider both traditional machine learning methods and
recent neural architectures. As a classical approach, we adopt Morgan
molecular fingerprints combined with a Random Forest (RF) regressor,
which is widely used for PCE prediction~\cite{eibeck2021predicting}. Among neural models, we
include a Transformer applied to molecular SMILES sequences, an MPNN
operating on molecular graphs, DropConn~\cite{10164235}, and two recent
graph transformer variants, RingFormer~\cite{ding2025ringformer} and
GRIT~\cite{ma2023graph}. DropConn is a graph neural network designed
for molecular property prediction using random connection dropping.
We also evaluate abcBERT~\cite{sun2024accelerating}, a model that
performs pretraining exclusively on the molecular graph modality.

We observe that many recent neural architectures, despite their success in other molecular tasks, underperform on PCE prediction compared to the traditional fingerprint-based approach. This suggests that relying on a single modality is often insufficient for capturing the complex structural and electronic factors that govern OSC efficiency. In particular, molecular fingerprints remain highly informative in this task, motivating their inclusion as an essential modality in our framework. Among baselines, abcBERT benefits from graph-level pretraining and shows competitive performance, yet it still lags behind our predictor. This shows the advantage of combining multimodal representations with uncertainty-aware learning in capturing the factors that determine PCE.

\subsubsection{Ablation Study}
The ablation study shows that molecular graphs, SMILES representations,
handcrafted molecular features, uncertainty quantification, and
pretraining each contribute to the predictor. In particular, removing
either the molecular graph branch or the SMILES branch leads to clear
performance degradation, indicating that the two modalities provide
complementary views of molecular structure, while handcrafted
descriptors remain highly informative in the limited-data OSC setting.
Overall, these components yield a more accurate and reliable predictor, providing OSCAgent with a solid foundation for downstream molecular design.

\begin{table}[t]
\centering
\caption{Performance comparison and ablation study for PCE prediction.}
\label{tab:pce_results}

\setlength{\tabcolsep}{6pt}
\renewcommand{\arraystretch}{1.15}

\resizebox{0.7\columnwidth}{!}{
\begin{tabular}{lcc}
\toprule
\textbf{Method} & \textbf{$R^2$} $\uparrow$ & \textbf{MAE} $\downarrow$ \\
\midrule
Morgan + RF~\cite{eibeck2021predicting} & 0.649 & 1.875 \\
Transformer~\cite{vaswani2017attention} & 0.554 & 2.128 \\
MPNN~\cite{gilmer2017neural} & 0.589 & 2.027 \\
DropConn~\cite{10164235} &  0.585&2.105 \\
RingFormer~\cite{ding2025ringformer} & 0.631 & 1.948 \\
GRIT~\cite{ma2023graph} & 0.643 & 1.901 \\
abcBERT~\cite{sun2024accelerating} & 0.668 & 1.781 \\
\midrule
\textbf{Ours (full)} & \textbf{0.713} & \textbf{1.686} \\
\midrule
w/o UQ & 0.681 & 1.776 \\
w/o Pretraining & 0.654 & 1.879 \\
w/o Handcrafted Feat & 0.634 & 1.924 \\
w/o Molecular Graph & 0.665 & 1.822 \\
w/o SMILES & 0.675 & 1.793 \\
\bottomrule
\end{tabular}
}
\end{table}

\subsubsection{Uncertainty Calibration Analysis}
\label{sec:uq_calibration}

\begin{figure}[!t]
\centering

\subfloat[]{%
    \includegraphics[width=0.48\columnwidth]
    {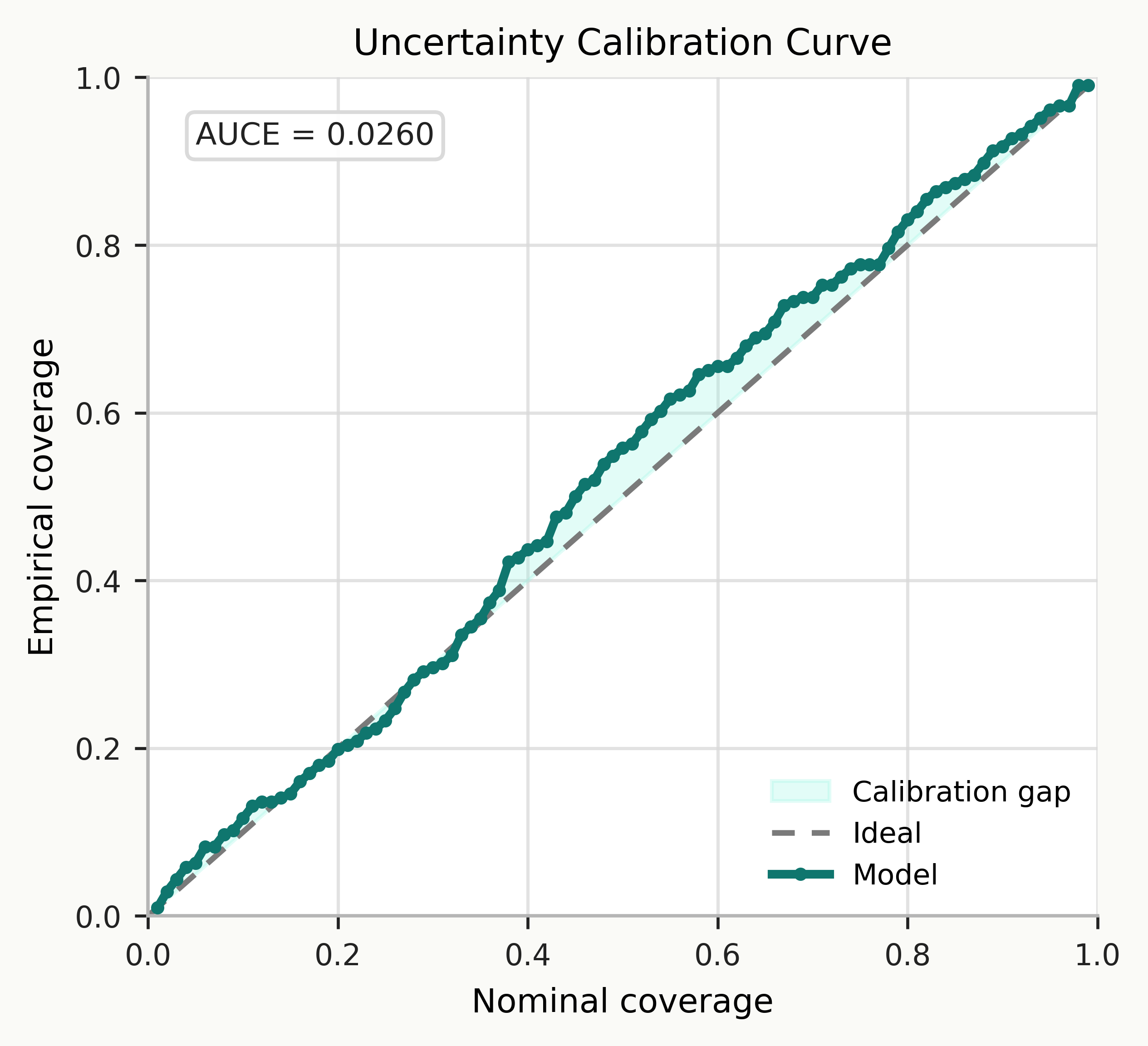}%
    \label{fig:uq_calibration_curve}
}
\hfill
\subfloat[]{%
    \includegraphics[width=0.48\columnwidth]
    {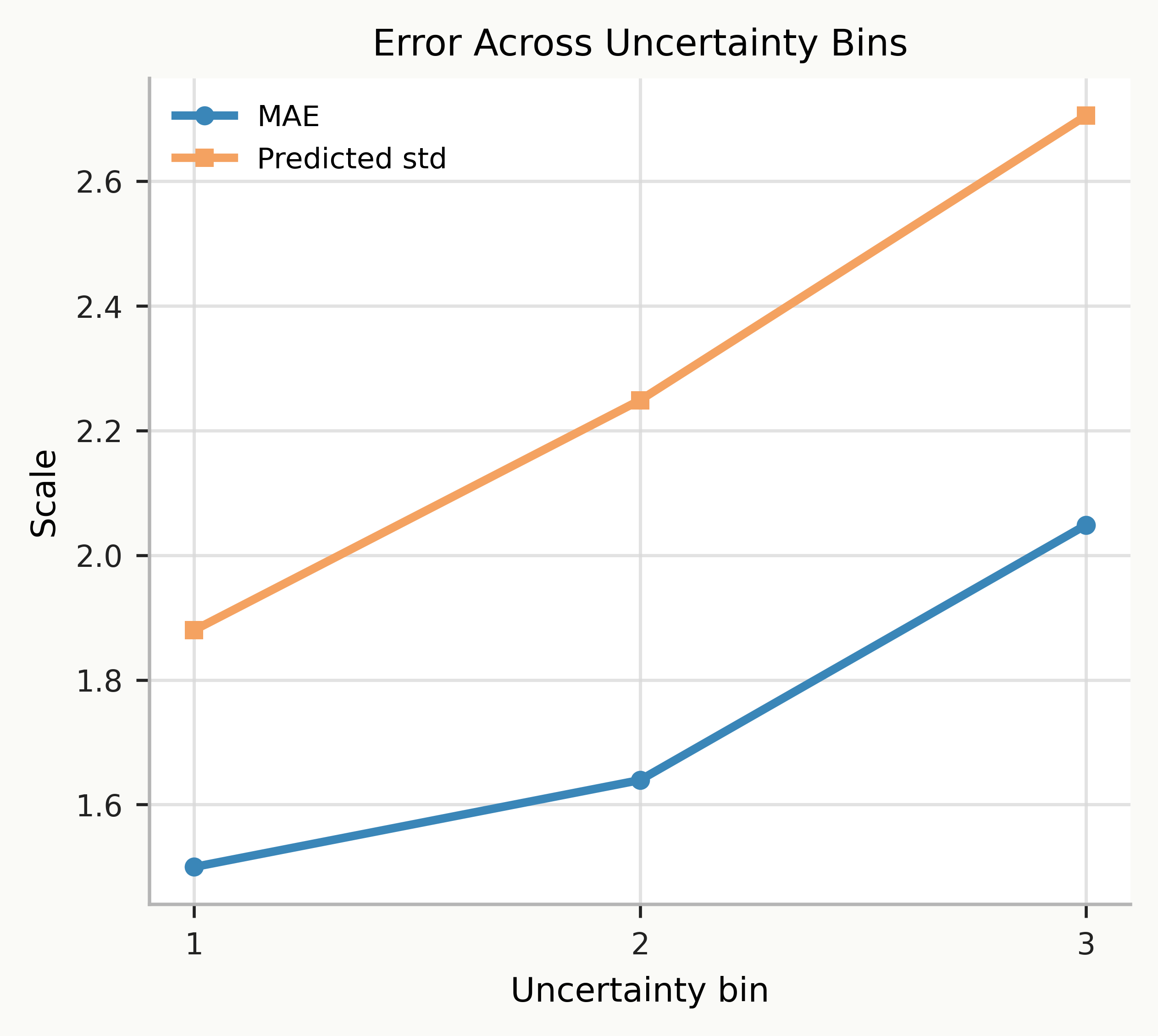}%
    \label{fig:uq_uncertainty_bins}
}

\vspace{1mm}

\subfloat[]{%
    \includegraphics[width=0.48\columnwidth]
    {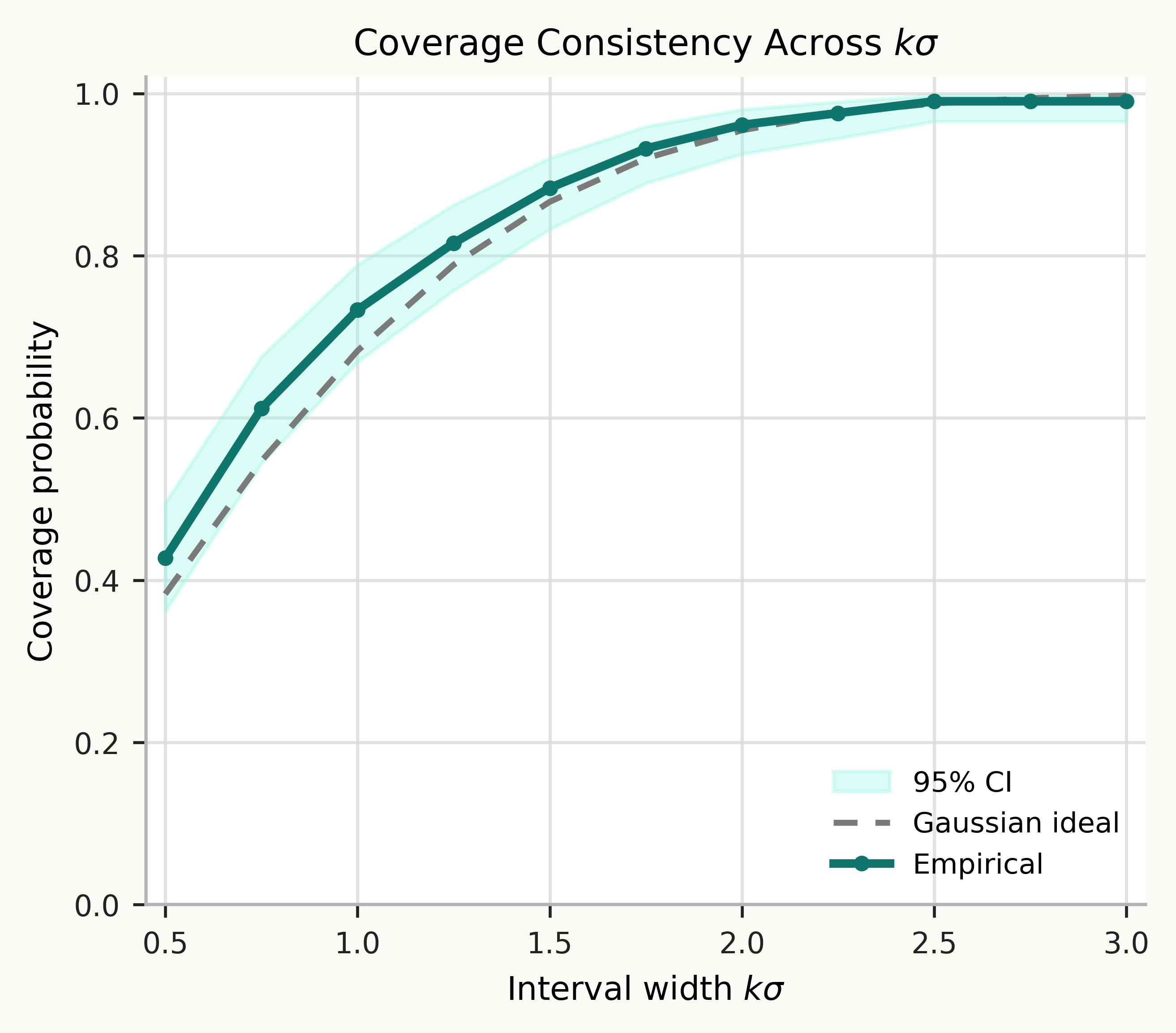}%
    \label{fig:uq_ksigma_coverage}
}
\hfill
\subfloat[]{%
    \includegraphics[width=0.48\columnwidth]
    {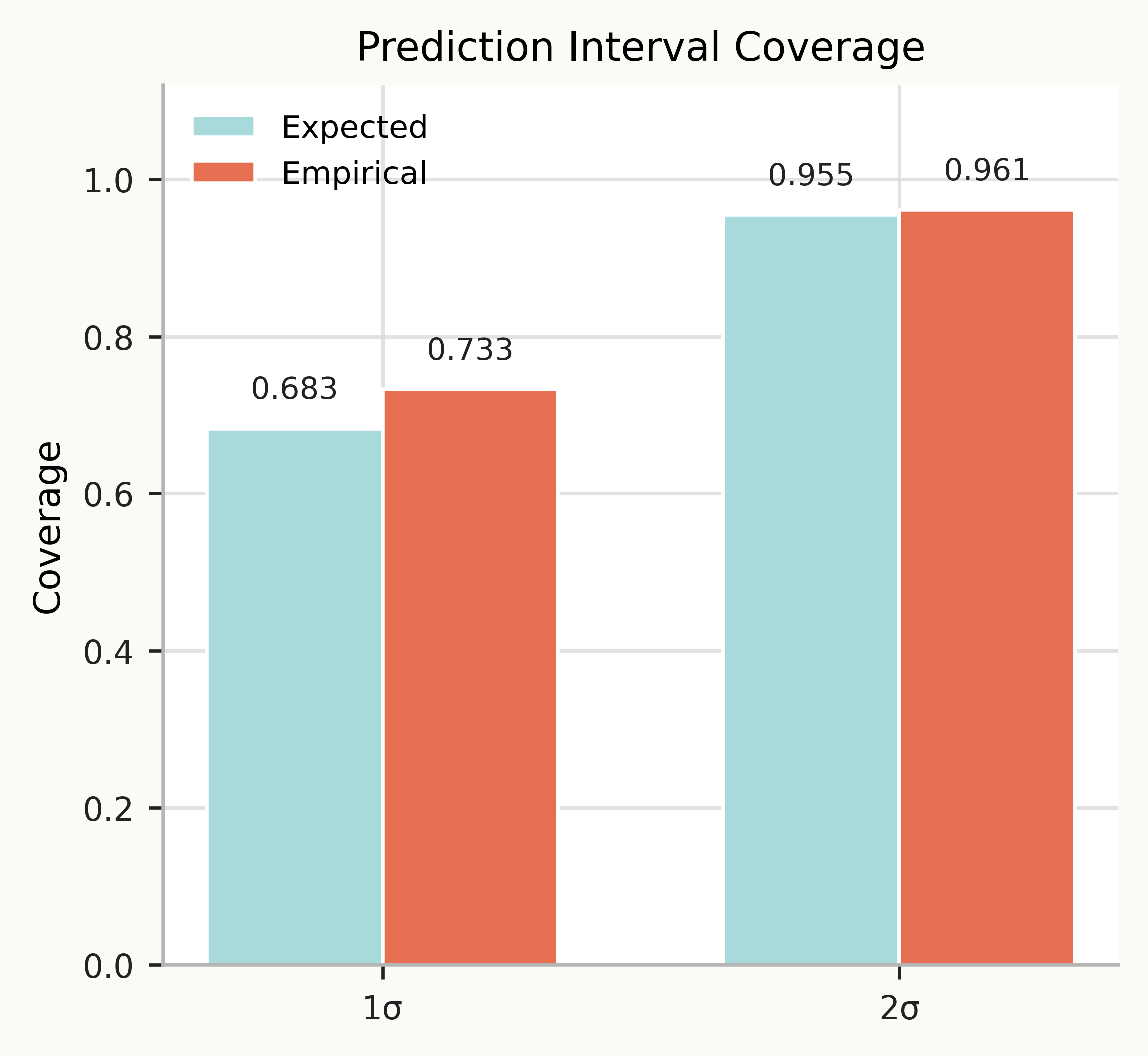}%
    \label{fig:uq_interval_coverage}
}

\caption{Uncertainty calibration of the proposed PCE predictor.
(a) Nominal versus empirical coverage.
(b) MAE and predicted uncertainty across bins.
(c) Coverage under different $k\sigma$ intervals.
(d) Expected and empirical coverage for the $1\sigma$ and $2\sigma$
intervals.}
\label{fig:uncertainty_calibration_results}
\end{figure}

We evaluate the calibration and error-awareness of the uncertainty
estimates produced by the PCE predictor, as summarized in
Fig.~\ref{fig:uncertainty_calibration_results}. The empirical coverage
closely follows the nominal coverage, yielding an area under the
calibration error curve (AUCE) of 0.026
(Fig.~\ref{fig:uncertainty_calibration_results}(a)). Moreover, samples
with larger predicted standard deviations exhibit higher empirical
MAE, indicating that the predicted uncertainty reflects the relative
difficulty and reliability of individual PCE predictions
(Fig.~\ref{fig:uncertainty_calibration_results}(b)).

We further evaluate prediction intervals of the form
\begin{equation}
\left[
\mu(\mathbf{x})-k\sigma(\mathbf{x}),
\,
\mu(\mathbf{x})+k\sigma(\mathbf{x})
\right],
\label{eq:pce_prediction_interval}
\end{equation}
where $\mu(\mathbf{x})$ and $\sigma(\mathbf{x})$ denote the predicted
PCE mean and standard deviation, respectively. The empirical coverage
follows the Gaussian reference trend as $k$ increases and remains
largely within the estimated 95\% confidence intervals
(Fig.~\ref{fig:uncertainty_calibration_results}(c)). In particular, the
empirical coverages of the $1\sigma$ and $2\sigma$ intervals are 0.733
and 0.961, respectively, compared with the corresponding Gaussian
reference values of 0.683 and 0.955
(Fig.~\ref{fig:uncertainty_calibration_results}(d)). These results
indicate that the predicted intervals are well calibrated, with a
slightly conservative tendency.

Within OSCAgent, the Experimenter reports both the predicted PCE and
its uncertainty to the Planner, allowing candidates with high predicted
performance but low confidence to be treated more cautiously.
Incorporating uncertainty into the agent feedback reduces the average
predicted uncertainty of generated molecules by 11.4\%, suggesting
that the design process is directed toward candidates with more
confident surrogate predictions.






\section{Conclusion}

In this work, we presented OSCAgent, a multi-agent framework for OSC
acceptor discovery that integrates retrieval-augmented planning, guided
molecular generation, comprehensive property evaluation, and
uncertainty-aware PCE prediction within a closed-loop workflow. Through
the collaboration of the Planner, Generator, and Experimenter,
OSCAgent iteratively refines its design strategies using knowledge from
experimentally validated molecules and feedback from previously
generated candidates. Experimental results demonstrate that OSCAgent
generates chemically valid, synthetically feasible, and high-performing
OSC acceptor candidates, outperforming traditional molecular generation
methods and LLM-based baselines across multiple evaluation metrics. The
multimodal PCE predictor further provides accurate predictions together
with uncertainty estimates, enabling more reliable feedback during
molecular exploration. Future work will extend the framework to broader
classes of functional materials and incorporate wet-lab experimental
feedback to achieve tighter integration between computational design
and experimental validation.

\ifCLASSOPTIONcaptionsoff
\newpage
\fi



%


\bibliographystyle{IEEEtran}
\bibliography{acm_mm_oscagent}
%




\end{document}